\documentclass[conference]{IEEEtran}
\IEEEoverridecommandlockouts
\ifCLASSINFOpdf
   \usepackage[pdftex]{graphicx}
   \graphicspath{{../pdf/}{../jpeg/}}
   \DeclareGraphicsExtensions{.pdf,.jpeg,.png}
\else
   \usepackage[dvips]{graphicx}
   \graphicspath{{../eps/}}
   \DeclareGraphicsExtensions{.eps}
\fi

\usepackage{fixltx2e}
\usepackage{amsmath,amssymb,amsfonts}
\usepackage{cite} 
\usepackage{textcomp}
\usepackage{algorithm}
\usepackage{multirow}
\usepackage{cuted}
\usepackage{bm}
\usepackage{float}
\usepackage{url}
\usepackage{diagbox}
\usepackage{graphics}
\usepackage{setspace}
\usepackage{threeparttable}
\usepackage[T1]{fontenc}
\usepackage{aecompl}
\usepackage{algorithm}
\usepackage{algpseudocode}  
\usepackage{amsthm}  

\newtheorem{proposition}{Proposition}
\usepackage{booktabs}

\usepackage{graphicx}
\usepackage[caption=false,font=footnotesize]{subfig}
\newtheorem{example}{Example}

\def\BibTeX{{\rm B\kern-.05em{\sc i\kern-.025em b}\kern-.08em
		T\kern-.1667em\lower.7ex\hbox{E}\kern-.125emX}}
\begin{document}

\title{An Ambiguity-Function-Assisted Newtonized Channel Estimation Method for Pulse-Shaped AFDM Under Fractional Delay and Doppler 
\thanks{$^{\ast}$ Corresponding author.}}


\author{\IEEEauthorblockN{Yuanfang Ma, Zulin Wang, Yuanhan Ni\IEEEauthorrefmark{1}, Tao Luo and Peng Yuan}
	\IEEEauthorblockA{School of Electronic and Information Engineering, Beihang University, Beijing, 100191, China\\
		Email: \{yuanfangma, wzulin, yuanhanni, luotao22,
		yuanpeng9208\}@buaa.edu.cn  \\
}}

\maketitle


\begin{abstract}

Accurate channel estimation for pulse-shaped AFDM systems over doubly selective channels with fractional normalized delay and Doppler remains challenging. This paper proposes a low-complexity ambiguity-function-assisted newtonized channel (AFNC) estimation method. Specifically, we first present a closed-form input-output relation for pulse-shaped affine frequency division multiplexing (AFDM) under fractional normalized delay and Doppler. 
As a further step, we demonstrate that the input-output relation admits a low-complexity representation by offline precomputing and storing the discretized ambiguity function of the shaping pulse, followed by tailored cyclic-shift and stacking operations.
Building on this representation, AFNC performs fractional delay-Doppler channel estimation through Newtonized refinement, where the required Jacobian and Hessian updates are computed efficiently using the low-complexity input-output representation. Simulation results confirm the effectiveness of the proposed approach.
\end{abstract}


%
\IEEEpeerreviewmaketitle

\section{Introduction}
Next-generation wireless communication systems (beyond 5G and toward 6G) are envisioned to support a wide range of services and applications, where providing high-throughput and reliable connectivity in high-mobility scenarios is a key requirement \cite{IMT2021White}. Such scenarios are often accompanied by severe Doppler spreads and rapidly time-varying multipath, which can significantly degrade link reliability, thus calling for robust and efficient physical-layer designs. As an enabling technology, waveform design is crucial for meeting this requirement.

Among various waveform candidates, orthogonal frequency-division multiplexing (OFDM) has become the de facto air-interface waveform and serves as the backbone of 5G New Radio (NR).
However, OFDM becomes increasingly vulnerable to Doppler-induced inter-carrier interference in high-mobility scenarios, motivating the exploration of alternative waveforms. \cite{Raviteja2018Interference,bemani2023affine}. Orthogonal time-frequency space (OTFS) spreads information symbols in the delay-Doppler (D-D) domain \cite{hadani2017orthogonal}, thereby handling large Doppler shifts and achieving full diversity in doubly selective channels. However, its two-dimensional modulation/demodulation typically incurs higher complexity, and its limited compatibility with OFDM-based transceivers may hinder deployment in predominantly OFDM-centric infrastructure. As an alternative chirp multicarrier waveform, AFDM multiplexes symbols in the affine domain via the discrete affine Fourier transform (DAFT) \cite{bemani2023affine,YTao2025IdxAFDM,Hyin2024MIMOAFDM,YNi2025ISAC,luo2024afdm}. This affine-domain formulation is well suited for doubly selective propagation and enables full diversity. Compared with OTFS, AFDM generally requires less pilot overhead and thus offers higher spectral efficiency. Moreover, with appropriate affine parameters, AFDM subsumes OFDM as a special case, making it attractive for practical OFDM-based systems.

Channel estimation is an indispensable component of AFDM communication systems. For AFDM without pulse shaping, channel estimation under integer-valued normalized delays and fractional normalized Dopplers has been investigated in prior works. In \cite{Cao2024ICCT}, orthogonal matching pursuit (OMP) exploits the delay-Doppler sparsity of doubly selective AFDM channels, yet it is sensitive to noise and highly coherent dictionaries. To enhance robustness, \cite{SBLtang} adopts sparse Bayesian learning (SBL), which is generally more resilient to noise and dictionary coherence, albeit at a substantially higher computational cost.
In practical AFDM systems, however, propagation delays are typically fractional with respect to the sampling grid, and pulse shaping (PS) is routinely employed for spectral confinement; both effects are therefore essential and must be accounted for. Nevertheless, this practically relevant regime has received limited attention, and systematic treatments are still lacking. Accordingly, it is necessary to develop channel-estimation methods for pulse-shaped AFDM under fractional delay and Doppler that jointly achieve high accuracy and low computational complexity.

This paper focuses on channel estimation for pulse-shaped AFDM under normalized fractional delay and Doppler. We first establish a closed-form input-output relation model.
By offline precomputing the discretized ambiguity function of the shaping pulse and then applying tailored cyclic-shift and stacking operations, we then obtain a low-complexity input-output representation.
Building on the proposed representation, we develop a low-complexity ambiguity-function-assisted Newtonized estimation (AFNC) method, where the Jacobian and Hessian matrix required for Newton refinement are efficiently updated using the low-complexity representation. 
Numerical results demonstrate that the proposed method achieves higher channel-estimation accuracy than SBL.
\section{Preliminaries}
\begin{figure}
	\centering
	\includegraphics[width=0.35\textwidth]{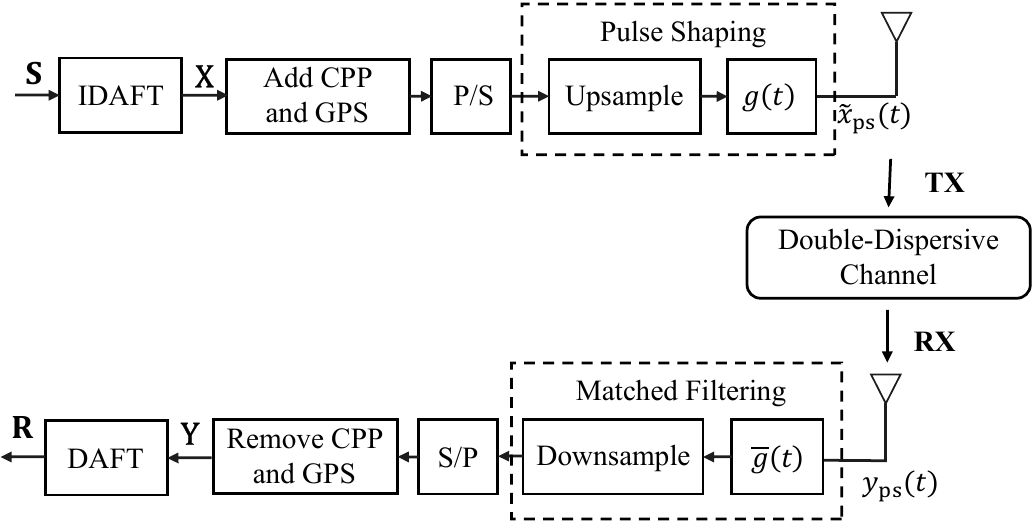}
	\caption{Block diagram of the AFDM transceiver with PS and matched filtering.}
	\label{fg:Block_diagram}
		\vspace{-15pt}
\end{figure}

In this section, we introduce the fundamental concept of AFDM, including PS at the transmitter and matched filtering at the receiver. Figure.~\ref{fg:Block_diagram} illustrates the overall system block diagram.


Let $\mathbf{S}[m,k]$ denote the $(m,k)$-th entry of an $N\times K$ affine-domain symbol block, where symbols are drawn from $\mathbb{A}$ (e.g., quadrature amplitude
modulation (QAM)), $m\in\{0,\ldots,N-1\}$ indexes subcarriers, and $k\in\{0,\ldots,K-1\}$ indexes symbols. Applying the inverse discrete affine Fourier transform (IDAFT) columnwise yields the time-domain matrix
\begin{equation} \label{eq: IDAFT}
\mathbf{X}[n,k] = \frac{1}{\sqrt{N}} e^{\mathrm{j} 2\pi c_1 n^2} \sum_{m=0}^{N-1} \mathbf{S}[m,k] \, e^{\mathrm{j} 2\pi \left( c_2 m^2 + \frac{m n}{N} \right)},
\end{equation}
where $n=0,1,...,N-1$. The corresponding matrix form of \eqref{eq: IDAFT} can be written as $\mathbf{X} = \mathbf{A}\mathbf{S}$ with ${\bf{A}} = {{\bf{\Lambda }}_{{c_2}}}{\bf{F}}_N{{\bf{\Lambda }}_{{c_1}}}$, $\mathbf{F}_N\in {\mathbb{C}^{N \times N}}$ being the discrete Fourier transform (DFT) matrix with entries ${e^{-{\rm{j}}2\pi pq/N}}/\sqrt N $ and 
\begin{equation} \label{eq: matrix_c}
\mathbf{\Lambda}_c = \operatorname{diag}\left(e^{-\mathrm{j} 2\pi c n^2},\; n=0,1,\dots,N-1\right).
\end{equation}
%
After modulation, a chirp-periodic prefix (CPP) of length $N_{\rm cp}$ is prepended to each column of $\mathbf{X}$, which is given by 
\begin{equation} \label{eq: CPP}
{{\bf{X}}_{{\rm{cp}}}}\left[ {n,k} \right] = {e^{ - \mathrm{j}2\pi {c_1}\left( {{N^2} + 2Nn} \right)}}{{\bf{X}}_{{\rm{cp}}}}\left[ {{{\left\langle n \right\rangle }_N},k} \right],
\end{equation}
where $N_{\rm cp}$ is chosen no smaller than the maximum channel delay spread (in samples) \cite{bemani2023affine}. When $2Nc_1$ is an integer and $N$ is even, the CPP reduces to a conventional periodic prefix (CP). To accommodate the pulse-shaping filter memory and avoid inter-symbol interference (ISI) across adjacent columns after P/S conversion, guard prefix and suffix (GPS) are further inserted for each column of $\mathbf{X}$. The guard prefix and suffix are given in \cite{GPS_Ni}, which are denoted as ${{\bf{X}}_{{\rm{gp}}}}$ and ${{\bf{X}}_{{\rm{gs}}}}$, respectively.
The GPS length per side (in samples) is $N_{\rm g}$. To accommodate the pulse-shaping filter memory, $N_{\rm g}$ is chosen to satisfy $N_{\rm g}\ge N_{\rm p}$, where $N_{\rm p}$ is the length (in samples) of the discretized pulse-shaping filter $g(t)$, which is defined in detail in Sec.~III-B.
After P/S conversion, define ${N_{{\rm{gps}}}} \buildrel \Delta \over = (N + {N_{{\rm{cp}}}} + 2{N_{\rm{g}}})K$, the discrete-time signal vector ${\widetilde {\bf{x}}_{{\rm{gps}}}} \in {{\mathbb{C}}^{{N_{{\rm{gps}}}} \times 1}}$ is given by
\begin{equation} \label{eq: x_gps}
{\widetilde {\bf{x}}_{{\rm{gps}}}} = {\mathop{\rm vec}\nolimits} \left( {{{\left[ {{\bf{X}}_{{\rm{gp}}}^{\rm{T}},\;{\bf{X}}_{{\rm{cp}}}^{\rm{T}},\;{{\bf{X}}^{\rm{T}}},\;{\bf{X}}_{{\rm{gs}}}^{\rm{T}}} \right]}^{\rm{T}}}} \right),
\end{equation}
with ${\rm{vec}}\left(  \cdot  \right)$ denoting the vectorization of matrix. Afterward, PS is applied to ${\widetilde {\bf{x}}_{{\rm{gps}}}}$. First, ${\widetilde {\bf{x}}_{{\rm{gps}}}}$ is upsampled with an upsampling factor $L$. The corresponding upsampled signal, denoted by ${\widetilde {\bf{x}}_{{\rm{gps,up}}}}\in {{\mathbb{C}}^{L{N_{{\rm{gps}}}} \times 1}}$, is given by
\begin{equation} \label{eq: up_sample}
{\widetilde {\bf{x}}_{{\rm{gps}},{\rm{up}}}} = {\left[ {{{\widetilde {\bf{x}}}_{{\rm{gps}},0}},\;{\bf{0}}_{L - 1}^{\rm{T}},\; \cdots ,\;{{\widetilde {\bf{x}}}_{{\rm{gps}},{N_{{\rm{gps}}}} - 1}},\;{\bf{0}}_{L - 1}^{\rm{T}}} \right]^{\rm{T}}}.
\end{equation}
The pulse-shaping filter $g\left( t \right)$ is effectively time-limited to $t \in \left[ { - {T_{\rm{p}}}/2,{T_{\rm{p}}}/2} \right]$ with duration $T_{\rm{p}}$, i.e., $g\left( t \right)$ is negligible outside this interval. 
${\widetilde {\bf{x}}_{{\rm{gps}},{\rm{up}}}}$ is linearly convolved with $g\left( t \right)$, yielding the pulse-shaped output ${\widetilde x_{{\rm{ps}}}}\left( t \right)$, which can be expressed 
\begin{equation} \label{eq: convolve}
{\widetilde x_{{\rm{ps}}}}\left( t \right) = \sum\limits_{n = 0}^{{N_{{\rm{gps}}}} - 1} {{{{\bf{\tilde x}}}_{{\rm{gps}},n}}} g\left( {t - nL{T_{\rm{s}}}} \right),
\end{equation}
where $T_{\rm{s}}$ denotes the sampling interval after upsampling, and thus the corresponding sampling rate is ${f_{\rm{s}}} = 1/{T_{\rm{s}}}$. Accordingly, ${\widetilde x_{{\rm{ps}}}}\left( t \right)$ is effectively time-limited to $t \in \left[ { - {T_{\rm{p}}}/2,{N_{{\rm{gps}}}}L{T_{\rm{s}}} + {T_{\rm{p}}}/2} \right]$. 
After propagation through the channel, the received signal can be expressed as
\begin{equation} \label{eq: receive signal}
\tilde y\left( t \right) = \sum\limits_{i = 1}^P {{h_i}} {e^{{\rm{j}}2\pi {{\tilde \nu}_i}t}}{\tilde x_{{\rm{ps}}}}(t - {\tilde \tau _i}) + w\left( t \right),
\end{equation}
where $h_i$, ${{{\tilde \nu}_i}}$ and ${\tilde \tau _i}$ denote the complex path gain, continuous-time Doppler shift and propagation delay of the $i$-th path, respectively. Here, $P$ is the number of channel paths and $w\left( t \right)$ denotes time-domain noise. Define ${\tau _{i}} = \frac{{{{\tilde \tau}_i}}}{{L{T_{\rm{s}}}}}$ and ${\nu_{i}} = {{{{\tilde \nu}_i}}}/\Delta f$, where $\Delta f = {f_{\rm{s}}}/\left( {NL} \right)$ denotes the subcarrier spacing. Hence, ${{{{\tau}_i}}}$ and ${{{{\nu}_i}}}$ represent fractional delay and Doppler normalized by ${{L{T_{\rm{s}}}}}$ and $\Delta f$, respectively.

At the receiver, ${y_{{\rm{ps}}}}\left( t \right)$ is first processed by a matched filter with impulse response $\bar g\left( t \right) = {g^*}\left( { - t} \right)$. The resulting signal is given by
\begin{equation} \label{eq: receiver matched filtering}
{\tilde y_{{\rm{mf}}}}\left( t \right) = \int {\tilde y\left( u \right)} {g^*}(u - t){\mkern 1mu} du.
\end{equation}
Subsequently, $\tilde y_{\rm mf}(t)$ is downsampled with interval $LT_{\rm s}$ to form
$\widetilde{\bf y}_{{\rm gps},n}=\tilde y_{\rm mf}(nLT_{\rm s})$ , where $n = 0, \ldots ,(N + {N_{{\rm{cp}}}} + 2{N_{\rm{g}}})K - 1$. The resulting sequence is reshaped via S/P into
$\overline{\bf Y}_{\rm gps}\in\mathbb{C}^{(N+N_{\rm cp}+2N_{\rm g})\times K}$. After removing the guard prefix/suffix and the CPP from each column, we obtain the received time-domain matrix ${\bf Y}\in\mathbb{C}^{N\times K}$. Applying the DAFT columnwise yields the affine-domain matrix ${\bf R}\in\mathbb{C}^{N\times K}$, given by
\begin{equation} \label{eq: R}
\mathbf{R}[p,k] = \frac{1}{\sqrt{N}} \sum_{n=0}^{N-1} \mathbf{Y}[n,k] 
e^{-\mathrm{j} 2\pi \left( c_1 n^2 + \frac{1}{N} pn + c_2 p^2 \right)}.
\end{equation}

\section{Low-Complexity Input-Output Representation Under Fractional Delay and Doppler}
In this section, we present a closed-form input-output relation for pulse-shaped AFDM under fractional normalized delay and Doppler, and further develop a low-complexity representation to reduce the computational complexity of the ensuing channel estimation algorithm. For notational simplicity, a single AFDM symbol (i.e., $K=1$) is considered in the following derivations, without loss of generality. Let ${\bf{s}}\left[ m \right] \buildrel \Delta \over = {\bf{S}}\left[ {m,1} \right]$ and ${\bf{r}}\left[ p \right] \buildrel \Delta \over = {\bf{R}}\left[ {p,1} \right]$.
%
\subsection{Closed-Form Input-Output Relation under Fractional Delay and Doppler}
\begin{proposition}
	The input-output relation for pulse-shaped AFDM under fractional delay and Doppler can be written as
	\begin{equation}\label{eq:io_prop}
	{\bf{r}} = \sum\limits_{i = 1}^P {{\beta _i}} {\bf{A}}{{\bf{\Delta }}_{{\nu _i}}}{{\bf{C}}_{{{\tilde \tau }_i},{{\tilde \nu }_i}}}{{\bf{A}}^{\rm{H}}}{\bf{s}} + {\bf{w}},
	\end{equation}
	where \(\mathbf{w}\) denotes the affine-domain noise and \(P\ge 1\) is the number of paths. The complex path gain ${\beta _i}$ is denoted as ${\beta _i} = {h_i}{e^{{\rm{j}}2\pi {{\tilde \nu}_i}{{\tilde \tau }_i}}}$.
	The Doppler-induced diagonal matrix ${{\bf{\Delta }}_{{{\nu}_i}}} \in {\mathbb{C}^{N \times N}}$ is defined as
	\begin{equation}\label{eq:Delta_fi}
	{{\bf{\Delta }}_{{\nu _i}}} \buildrel \Delta \over = {\rm{diag}}\left( {\left\{ {{e^{{\rm{j}}2\pi {\nu _i}n/N}}} \right\}_{n = 0}^{N - 1}} \right).
	\end{equation}
	The \((p,m)\)-th entry of \(\mathbf{C}_i\), for \(p,m\in\{0,1,\ldots,N-1\}\), is given by
	\begin{equation}\label{eq:Cim}
	{{\bf{C}}_{{{\tilde \tau }_i},{{\tilde \nu }_i}}}[p,m] = \sum\limits_{q \in \mathbb{Z}} {{\varphi _i}^{ - l}} {\mkern 1mu} {A_g}\left( {lL{T_{\rm{s}}} - {{\tilde \tau }_i},\;{{{\tilde \nu}_i}}} \right),
	\end{equation}
	where \(\varphi_i \triangleq e^{\mathrm{j}2\pi {{{\nu}_i}} /N}\) and $l \buildrel \Delta \over = p - m - qN$. Moreover, \(A_g(\tau,\nu)\) denotes the ambiguity function of \(g(t)\), defined as
	\begin{equation}\label{eq:ambiguity_g}
	A_g(\tau,\nu)
	=\int_{-\infty}^{\infty} g(t)\, g^{*}(t-\tau)\, e^{\mathrm{j}2\pi \nu t}\, \mathrm{d}t,
	\end{equation}
	and \(g^{*}(t)\) denotes the complex conjugate of \(g(t)\).
	
	\textit{Proof:} Due to space limitations, the proof will be provided in future work.
\end{proposition}

With the closed-form input-output relation, the received signal in the affine domain can be evaluated directly for arbitrary ${\bf{s}}$ and $\left( {{{\tilde \tau }_i},{{\tilde \nu }_i}} \right)$. Directly evaluating \eqref{eq:io_prop} is computationally prohibitive, since forming ${\bf C}_i$ requires evaluations of $A_g(\tau,\nu)$ (which requires an integral) and the subsequent dense matrix operations lead to an overall complexity on the order of $O\left( {P{N^3}} \right)$. Therefore, we develop a low-complexity input--output representation in Sec.~III-B.

\subsection{Low-Complexity Input-Output Representation}
The key idea is to precompute and store the discretized samples of $A_g(\tau,\nu)$ offline, and obtain the low-complexity computation of ${{\bf{C}}_{{{\tilde \tau }_i},{{\tilde \nu }_i}}}$ via aligned cyclic shifts, which can be efficiently implemented using Fast Fourier Transform (FFT)-based operations. 
Moreover, a stacking operation is employed to provide a low-complexity representation of the delay- and Doppler-induced distortion on the symbols in the affine domain, thereby avoiding explicit multiplications by large structured matrices. The low-complexity representation of the $i$-th-path affine-domain received signal is given by \eqref{eq:In_out_fraCase} at the top of next page. Then, the overall affine-domain received signal can be expressed as ${\bf{r}} = \sum\limits_{i = 1}^P {{\beta _i}{{\bf{r}}_i}}$. For brevity, the matrix definitions are deferred to the following discussion.
\begin{figure*}[ht]
	\vspace{-6pt}
	\begin{align}\label{eq:In_out_fraCase}		
	{{\bf{r}}_i} = \sum\limits_{{n_{\rm{c}}} = 0}^{N - 1} {\bf{s}}\left[ {{n_{\rm{c}}}} \right]\left( {{{\bf{F}}_N}{{\bf{\Delta }}_{{{\nu}_i}}}{{\bf{F}}_N}^{\rm{H}}} \right){{{\bf{\Pi }}^{{n_{\rm{c}}}}}{{\bf{T}}_{{n_{\rm{c}}}}}{{\bf{\Psi }}_{{{\nu}_i}}}{{\bf{D}}_L}\left( {{{\bf{F}}_{NL}}{{\bf{\Lambda }}_{{\tau _i}}}{\bf{F}}_{NL}^{\rm{H}}} \right){{\bf{\Pi }}_{{\rm{sh}}}}{\bf{P}}{{{\bf{\bar A}}}_g}\left( {{{\bf{F}}_{NL}}{{\bf{\Lambda }}_{{\nu _i}}}{\bf{F}}_{NL}^{\rm{H}}} \right){{\bf{e}}_{{u_0}}}}.
	\end{align}
	\vspace{-5pt}
	\hrule
\end{figure*}
\subsubsection{Low-Complexity Computation of ${{\bf{C}}_{{{\tilde \tau }_i},{{\tilde \nu }_i}}}$}


Let ${\bf g}\in\mathbb{C}^{N_{\rm p}\times 1}$ denote the discretized version of $g(t)$ with sampling interval $T_{\rm s}$, whose $n$-th entry is
${\bf g}[n]=g\!\big((n-N_{\rm p}/2)T_{\rm s}\big)$, where $N_{\rm p}=T_{\rm p}/T_{\rm s}$ and $n\in\{0,\ldots,N_{\rm p}-1\}$. We precompute and store the sampled ambiguity function in ${\bf \bar A}_g\in\mathbb{C}^{(2N_{\rm p}+1)\times NL}$, whose $(p,u)$-th entry is given by
\begin{equation}\label{eq:bar_Ag1}
{{\bf{\bar A}}_g}\left[ {p,u} \right] \buildrel \Delta \over = {A_g}\left( {\left( {p - {N_{\rm{p}}}} \right){T_{\rm{s}}},\left( {u - \frac{{NL}}{2}} \right)\Delta f} \right),
\end{equation}
with $p \in \left\{ {0,1, \ldots ,2{N_{\rm{p}}}} \right\}$ and $u \in \left\{ {0,1, \ldots ,NL - 1} \right\}$. Using the discrete-time pulse ${\bf{g}}$, we have 
\begin{equation}\label{eq:bar_Ag2}
{{\bf{\bar A}}_g}\left[ {p,u} \right] = {T_{\rm{s}}}\sum\limits_{n = 0}^{N - 1} {{\bf{g}}\left[ n \right]{{\bf{g}}^*}\left[ {n - \left( {p - {N_{\rm{p}}}} \right)} \right]{e^{{\rm{j}}2\pi \left( {\frac{u}{{NL}} - \frac{1}{2}} \right)n}}}.
\end{equation}
Note that ${\bf{g}}[n]$ has finite support, i.e., ${\bf{g}}[n] = 0$ for $n \notin \{ 0, \ldots ,{N_{\rm{p}}} - 1\} $. 
In \eqref{eq:In_out_fraCase}, $\big({\bf F}_{NL}{\bf \Lambda}_{\nu_i}{\bf F}_{NL}^{\rm H}\big)\in\mathbb{C}^{NL\times NL}$ implements an FFT-based fractional circular shift along the Doppler axis of ${\bf \bar A}_g$ to align the stored Doppler bins with $\nu_i$, where
${\bf \Lambda}_{\nu_i}\triangleq{\rm diag}\!\left(e^{{\rm j}2\pi \nu_i n/NL}\right)$ with $n \in \left[ {0,0.5NL - 1} \right] \cup \left[ { - 0.5NL, - 1} \right]$, and ${\bf F}_{NL}$ is the $NL$-point DFT matrix. The vector ${\bf e}_{u_0}\in\mathbb{C}^{NL\times 1}$ is the $u_0$-th canonical basis vector with $u_0\triangleq NL/2$, used to extract the centered Doppler bin after the shift. The embedding matrix ${\bf P}\in\mathbb{C}^{NL\times(2N_{\rm p}+1)}$ then zero-pads the extracted delay slice (of length $2N_{\rm p}+1<NL$) into a length-$NL$ vector by placing it around the center index. Specifically,
\begin{equation}\label{eq:P_matrix}
{\bf{P}}[m,r] = \left\{ {\begin{array}{*{20}{l}}
	{1,}&{m = {m_0} + r}\\
	{0,}&{{\rm{otherwise}}}
	\end{array}} \right.,\quad r = 0, \ldots ,2{N_p},
\end{equation}
with ${m_0} = NL/2 - {N_{\rm{p}}}$. The permutation matrix ${{\bf{\Pi }}_{{\rm{sh}}}} \buildrel \Delta \over = {{\bf{\Pi }}^{NL/2}}\in {\mathbb{C}^{NL \times NL}}$ swaps the two length-$NL/2$ halves, where ${\bf{\Pi }}$ is the one-step forward cyclic-shift matrix \cite{bemani2023affine}. 
Next, ${\left( {{{\bf{F}}_{NL}}{{\bf{\Lambda }}_{{\tau _i}}}{\bf{F}}_{NL}^{\rm{H}}} \right)}\in {\mathbb{C}^{NL \times NL}}$ applies a fractional circular shift to the length-$NL$ vector, thereby realizing the normalized delay ${\tau _{i}}$, with ${{\bf{\Lambda }}_{{\tau _i}}} \buildrel \Delta \over = {\rm{diag}}\left( {{e^{ - {\rm{j}}2\pi {\tau _i}n/N}}} \right)$, and $n \in \left[ {0,0.5NL - 1} \right] \cup \left[ { - 0.5NL, - 1} \right]$. The matrix ${{\bf{D}}_L} \in {\mathbb{R}^{N \times NL}}$ then downsamples the length-$NL$ sequence by a factor of 
$L$ to match the delay-sampling interval $LT_{\rm{s}}$ in \eqref{eq:Cim}, i.e.,
\begin{equation}\label{eq:DL}
{{\bf{D}}_L}[n,m] = \left\{ {\begin{array}{*{20}{l}}
	{1,}&{m = nL}\\
	{0,}&{{\rm{otherwise}}}
	\end{array}} \right.,\quad n = 0, \ldots ,N - 1.
\end{equation}
Then, ${{\bf{\Psi }}_{{{\nu}_i}}}$ is defined as ${{\bf{\Psi }}_{{{\nu}_i}}} \buildrel \Delta \over = {\bf{\Delta }}_{{\nu _i}}^{ - 1}{{\bf{\Gamma }}_i}$. The diagonal matrix ${{\bf{\Gamma }}_i} \buildrel \Delta \over = {\mathop{\rm diag}\nolimits} \left( {{\gamma _i}[0], \ldots ,{\gamma _i}[N - 1]} \right)$, where ${\gamma _i}[m]$ is given by 
\begin{equation}\label{eq:Gama_matrix}
{\gamma _i}[m] = \left\{ {\begin{array}{*{20}{l}}
	{{\varphi _i}^N,}&{m \in \left\{ {N/2, \ldots ,N - 1} \right\},}\\
	{1,}&{{\rm{otherwise}}.}
	\end{array}} \right.
\end{equation}
The matrix ${{\bf{\Psi }}_{{{\nu}_i}}}$ applies the residual linear-phase rotation across the downsampled index $n$, which arises from the factor ${{\varphi _i}^{ - l}}$ in \eqref{eq:Cim}. By this point, we have obtained ${{\bf{c}}_{i,0}} \in {\mathbb{C}^{N \times 1}}$, which denotes the first column of ${{\bf{C}}_i}$. It can be expressed as 
\begin{equation}\label{eq:ci}
{{\bf{c}}_{i,0}} = {{\bf{\Psi }}_{{{\nu}_i}}}{{\bf{D}}_L}\left( {{{\bf{F}}_{NL}}{{\bf{\Lambda }}_{{\tau _i}}}{\bf{F}}_{NL}^{\rm{H}}} \right){{\bf{\Pi }}_{{\rm{sh}}}}{\bf{P}}{{\bf{\bar A}}_g}\left( {{{\bf{F}}_{NL}}{{\bf{\Lambda }}_{{\nu _i}}}{\bf{F}}_{NL}^{\rm{H}}} \right){{\bf{e}}_{{u_0}}}.
\end{equation}
Noting that ${{\bf{C}}_{{{\tilde \tau }_i},{{\tilde \nu }_i}}}$ in \eqref{eq:Cim} is a circulant matrix, it can be decomposed into a linear combination of forward cyclic-shift matrix as
\begin{equation}\label{eq:Cim_combination}
{{\bf{C}}_{{{\tilde \tau }_i},{{\tilde \nu }_i}}} = \sum\limits_{{i^\prime } = 0}^{N - 1} {{{\bf{c}}_{i,0}}\left[ {{i^\prime }} \right]{{\bf{\Pi }}^{{i^\prime }}}}.
\end{equation}
\subsubsection{Stacking Operation}
\begin{figure}
	\centering
	\includegraphics[width=0.3\textwidth]{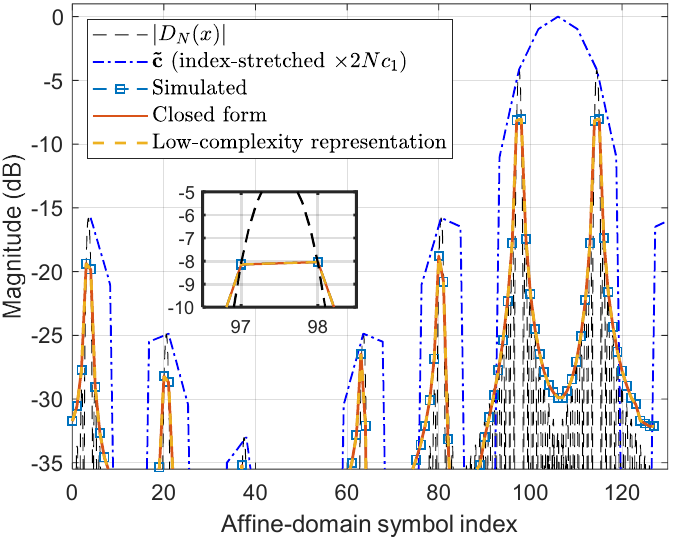}
	\caption{Magnitude (dB) of the affine-domain received pilot signal under fractional normalized delay and Doppler: simulation versus the closed-form model and the proposed low-complexity representation.}
	\vspace{-15pt}
	\label{fg:received_signal}
\end{figure}

By applying a stacking operation to ${{\bf{c}}_{i,0}}$, the affine-domain received signal can be obtained with low complexity.
Owing to the circulant structure of ${{\bf{C}}_{{{\tilde \tau }_i},{{\tilde \nu }_i}}}$, a single-path fractional-delay channel can be equivalently represented as a superposition of integer-delay components in the affine-domain equivalent representation, where the ${i^\prime}$-th component has complex gain ${\bf c}_{i,0}[i^\prime]$. Since each integer-delay component induces a distinct cyclic shift of $2Nc_1$ in the affine domain, the resulting superposition can be written as
\begin{equation}\label{eq:Stack_c}
{{\bf{r}}_{{\rm{d}},i}} = \sum\limits_{{n_{\rm{c}}} = 0}^{N - 1} {\bf{s}}\left[ {{n_{\rm{c}}}} \right]{{\bf{\Pi }}^{{n_{\rm{c}}}}}{{\bf{T}}_{{n_{\rm{c}}}}}{{\bf{c}}_{i,0}},
\end{equation}
with ${n_{\rm{c}}}$ denoting the index of the affine-domain transmitted symbol ${\bf{s}}$, and ${{\bf{T}}_{{n_{\rm{c}}}}} \in {\mathbb{C}^{N \times N}}$ is constructed as
\begin{equation}\label{eq:Stack}
{{\bf{T}}_{{n_{\rm{c}}}}} = {{\bf{\Lambda }}_{{n_{\rm{c}}}}}{\bf{\bar \Lambda }}\left[ {\begin{array}{*{20}{l}}
	{{{\bf{e}}_{{n_0}}}}&{{{\bf{e}}_{{n_1}}}}& \cdots &{{{\bf{e}}_{{n_{N - 1}}}}}
	\end{array}} \right],
\end{equation}
with 
\begin{equation}\label{eq:n_k}
{n_k} \buildrel \Delta \over = {\langle N - 2Nc_1k\rangle _N} = {\langle  - 2Nc_1k\rangle _N},
\end{equation}
where $k=0,1,\ldots,N-1$, ${\bf e}_{n_k}\in\mathbb{C}^{N\times 1}$ denotes the $n_k$-th canonical basis vector, and $\langle\cdot\rangle_N$ is the modulo-$N$ operation. We define ${\bf{\bar \Lambda }} \buildrel \Delta \over = {\rm{diag}}\left( {\left\{ {{e^{{\rm{j}}2\pi {c_1}{k^2}}}} \right\}_{k = 0}^{N - 1}} \right)$. For subcarrier index $n_{\rm c}$, the corresponding phase-rotation matrix is ${{\bf{\Lambda }}_{{n_{\rm{c}}}}} \buildrel \Delta \over = {\rm{diag}}\left( {\left\{ {{e^{ - {\rm{j}}2\pi k{n_{\rm{c}}}/N}}} \right\}_{k = 0}^{N - 1}} \right)$. Fractional Doppler spreads the energy of ${\bf r}_{{\rm d},i}$ across multiple symbol indices. Equivalently, the resulting affine-domain symbol can be expressed as a convolution with the length-$N$ Dirichlet kernel, i.e.,
\begin{equation}\label{eq:Dirichlet}
{{\bf{r}}_i}[p] = \sum\limits_{n = 0}^{N - 1} {{{\bf{r}}_{{\rm{d}},i}}} [n]{D_N}\left( {p - n + {\nu _i}} \right),
\end{equation}
where 
\begin{equation}\label{eq:Dirichlet_kernal}
{D_N}(x) \buildrel \Delta \over = \frac{{{e^{ - {\rm{j}}2\pi x}} - 1}}{{{e^{ - {\rm{j}}\frac{{2\pi }}{N}x}} - 1}}.
\end{equation}
By computing the convolution via the DFT matrix, the resulting affine-domain received signal can be expressed as
\begin{equation}\label{eq:Dirichlet_DFT}
{{\bf{r}}_i} = {{\bf{F}}_N}{{\bf{\Delta }}_{{{\nu}_i}}}{\bf{F}}_N^{\rm{H}}{{\bf{r}}_{{\rm{d}},i}}.
\end{equation}
Combining \eqref{eq:Dirichlet_DFT}, \eqref{eq:Stack_c}, and \eqref{eq:ci} yields \eqref{eq:In_out_fraCase}. Owing to space constraints, a detailed and rigorous derivation of \eqref{eq:In_out_fraCase} will be provided in future work. When an affine-domain pilot is transmitted, the $i$-th-path received signal can be expressed as 
\begin{equation}\label{eq:ri_pilot}
{{\bf{r}}_i} = {\bf{s}}\left[ 0 \right]\left( {{{\bf{F}}_N}{{\bf{\Delta }}_{{{\nu}_i}}}{{\bf{F}}_N}^{\rm{H}}} \right){{\bf{T}}_0}{{\bf{c}}_{i,0}}.
\end{equation}
\begin{example}
We consider a single-path channel ($P = 1$), employ a root-raised-cosine (RRC) pulse-shaping filter $g\left( t \right)$, set $L = 4$, $N = 128$, $2Nc_1 = 17$ and $c_2 = 0$. The normalized delay and Doppler are chosen as ${{{{\tau}_0}}}=1.5$ and ${{{{\nu}_0}}}=3.5$, respectively. In Fig.~\ref{fg:received_signal}, we plot the affine-domain received pilot signal of a pulse-shaped AFDM system under fractional normalized delay and Doppler with an affine-domain pilot transmitted. The curves labeled ``Simulated'', ``Closed form'', and ``Low-complexity representation'' correspond to the simulation result, the derived closed-form model, and the proposed low-complexity model, respectively. The three curves overlap, validating the correctness of both analytical representations. Let ${{\bf{\tilde c}}}\in {\mathbb{C}^{NL \times 1}}$ be defined as
\begin{equation}\label{eq:ci_upsamp}
{{\bf{\tilde c}}} \buildrel \Delta \over = \left( {{{\bf{F}}_{NL}}{{\bf{\Lambda }}_{{\tau _0}}}{\bf{F}}_{NL}^{\rm{H}}} \right){{\bf{\Pi }}_{{\rm{sh}}}}{\bf{P}}{{\bf{\bar A}}_g}\left( {{{\bf{F}}_{NL}}{{\bf{\Lambda }}_{{\nu _0}}}{\bf{F}}_{NL}^{\rm{H}}} \right){{\bf{e}}_{{u_0}}},
\end{equation}
which represents the (pre-downsampling) delay slice extracted from ${{\bf{\bar A}}_g}$ after applying the Doppler- and delay-domain circular shifts. The curve labeled ``${{\bf{\tilde c}}}$ (index-stretched $ \times 2N{c_1}$)'' plots the envelope of ${{\bf{\tilde c}}}$ after index stretching by $2Nc_1$. The curve labeled ``$|D_N(x)|$'' shows the envelope of the length-$N$ Dirichlet kernel. It can be observed that a single-path channel with fractional delay can be decomposed into multiple integer-delay components whose amplitude envelope is given by ${{\bf{\tilde c}}}$, and the energy of each component is further spread according to the Dirichlet-kernel envelope $|D_N(x)|$.
\end{example}
\section{Low-Complexity AFNC Method For Channel With Fractional Normalized Delay and Doppler}
This section proposes a low-complexity AFNC channel estimation method for pulse-shaped AFDM under fractional normalized delay and Doppler. It builds on the low-complexity received-signal expression in \eqref{eq:ri_pilot} and a Newtonized refinement procedure.
\subsection{Low-Complexity AFNC Channel Estimation Method}
Let ${\bf r}_{\rm prac}$ denote the practical affine-domain received signal, and let $(\hat\beta_i,\hat\tau_i,\hat\nu_i)$ denote the estimates of the complex gain, normalized delay, and normalized Doppler of the $i$-th path, respectively. The proposed AFNC method consists of three stages: i) coarse estimation, ii) single refinement, and iii) cyclic refinement.

\subsubsection{Coarse-Estimation Stage}
In the initialization stage, we set the signal residual as ${{\bf{h}}_{\rm{r}}} = {{\bf{r}}_{{\rm{prac}}}}$. 
We then locate the peak of ${\bf h}_{\rm r}$, whose index is given by
\begin{equation}
p_0 \triangleq \arg\max_{p\in\{0,1,\ldots,N-1\}} \left|{\bf h}_{\rm r}[p]\right|.
\end{equation}
Exploiting the affine-domain cyclic-shift structure induced by integer normalized delay and Doppler, the integer normalized delay is estimated as
\begin{equation}\label{eq:Coarse_tau}
{\hat \tau _{i}} = {\rm{round}}\left( {\frac{{N - {p_0}}}{{2N{c_1}}}} \right),
\end{equation}
where ${\rm{round}}\left(  \cdot  \right)$ rounds to the nearest integer. And the corresponding integer normalized Doppler estimate is given by
\begin{equation}\label{eq:Coarse_Doppler}
{{\hat \nu}_{i}} = {p_0} - N + 2N{c_1}{\hat \tau _{i}}.
\end{equation}
\subsubsection{Single-Refinement Stage}
The refined estimates are obtained by solving the nonlinear least squares problem $\left\| {{{\bf{h}}_{\rm{r}}} - {{\hat \beta }_i}{{\bf{r}}_i}({{\hat \tau }_{i}},{{\hat \nu}_{i}})} \right\|_2^2$, which is equivalent to maximizing
\begin{equation}\label{eq:S_func}
S = 2\Re \left\{ {{\bf{h}}_r^{\rm{H}}{{\bf{r}}_i}({{\hat \tau }_{i}},{{\hat \nu}_{i}}){{\hat \beta }_i}} \right\} - {\left| {{{\hat \beta }_i}} \right|^2}\left\| {{{\bf{r}}_i}({{\hat \tau }_{i}},{{\hat \nu}_{i}})} \right\|_2^2.
\end{equation}
We update ${({{\hat \tau }_{i}},{{\hat \nu}_{i}})}$ using Newton iterations given by
\begin{equation}\label{eq:Newton_iter}
\left[ {\begin{array}{*{20}{c}}
	{{{\hat \tau }_{i}}}\\
	{{{\hat \nu}_{i}}}
	\end{array}} \right] = \left[ {\begin{array}{*{20}{c}}
	{{{\hat \tau }_{i}}}\\
	{{{\hat \nu}_{i}}}
	\end{array}} \right] - \mathop {\bf{S}}\limits^{..} {\left( {{{\hat \beta }_i},{{\hat \tau }_{i}},{{\hat \nu}_{i}}} \right)^{ - 1}}\mathop {\bf{S}}\limits^. \left( {{{\hat \beta }_i},{{\hat \tau }_{i}},{{\hat \nu}_{i}}} \right),
\end{equation}
where $\mathop {\bf{S}}\limits^.$ and $\mathop {\bf{S}}\limits^{..}$ denote the Jacobian matrix and the Hessian matrix, respectively, given by 
\begin{equation}\label{eq:Jacobian}
\mathop {\bf{S}}\limits^. \left( {{{\hat \beta }_i},{{\hat \tau }_{i}},{{\hat \nu}_{i}}} \right) = \left[ {\begin{array}{*{20}{c}}
	{\frac{{\partial S}}{{\partial {{\hat \tau }_{i}}}}}\\
	{\frac{{\partial S}}{{\partial {{\hat \nu}_{i}}}}}
	\end{array}} \right],
\end{equation}
\begin{equation}\label{eq:Hessian}
\mathop {\bf{S}}\limits^{..} \left( {{{\hat \beta }_i},{{\hat \tau }_{i}},{{\hat \nu}_{i}}} \right) = \left[ {\begin{array}{*{20}{c}}
	{\frac{{{\partial ^2}S}}{{\partial {{\hat \tau }^2}_{i}}}}&{\frac{{{\partial ^2}S}}{{\partial {{\hat \tau }_{i}}\partial {{\hat \nu}_{i}}}}}\\
	{\frac{{{\partial ^2}S}}{{\partial {{\hat \nu}_{i}}\partial {{\hat \tau }_{i}}}}}&{\frac{{{\partial ^2}S}}{{\partial {{\hat \nu}^2}_{i}}}}
	\end{array}} \right].
\end{equation}
The first- and second-order partial derivatives of $S$ with respect to ${{{\hat \tau }_{i}}}$ and ${{{\hat \nu}_{i}}}$ can be obtained from (18)-(22) in \cite{NOMP-OFDM} by substituting $({\hat \beta _l},{\hat \tau _l},{\hat \alpha _l},{\bf{a}})$ with $({\hat \beta _i},{\hat \tau _{i}},{\hat {{{\nu}_i}}},{{\bf{r}}_i})$. The main challenge then lies in evaluating $\partial {{\bf{r}}_i}/\partial {\hat \tau _{i}}$, $\partial {{\bf{r}}_i}/\partial {\hat {{{\nu}_i}}}$, as well as the corresponding second-order derivatives, which are required for calculating the derivatives of $S$. To facilitate the derivation of the partial derivatives of ${{\bf{r}}_i}$, we rewrite ${{\bf{r}}_i}$ in \eqref{eq:ri_pilot} in the following factorized form:
\begin{equation}\label{eq:r_factorized}
{{\bf{r}}_i} = {{\bf{M}}_1}{{\bf{\Delta }}_{{{\hat \nu}_i}}}{{\bf{M}}_2}{{\bf{\Psi }}_{{{\hat \nu}_i}}}{{\bf{M}}_3}{{\bf{\Lambda }}_{{\hat \tau _{i}}}}{{\bf{M}}_4}{{\bf{\Lambda }}_{{\hat \nu_{i}}}}{\bf{m}}.
\end{equation}
Here, ${{\bf{M}}_1} ={\bf{s}}\left[ 0 \right]{{\bf{F}}_N}$, ${{\bf{M}}_2} = {{\bf{F}}_N}^{\rm{H}}{\bf{T}}_0$, ${{\bf{M}}_3} = {{\bf{D}}_L}{{\bf{F}}_{NL}}$, ${{\bf{M}}_4} = {\bf{F}}_{NL}^{\rm{H}}{{\bf{\Pi }}_{{\rm{sh}}}}{\bf{P}}{{\bf{\bar A}}_g}{{\bf{F}}_{NL}}$ and ${\bf{m}} = {\bf{F}}_{NL}^{\rm{H}}{{\bf{e}}_{{u_0}}}$. All matrices $\left\{ {{{\bf{M}}_1},{{\bf{M}}_2},{{\bf{M}}_3},{{\bf{M}}_4}} \right\}$ and the vector ${\bf{m}}$ are constant and independent of  ${{\hat \tau _{i}}}$ and ${\hat \nu_{i}}$. Therefore, $\partial {{\bf{r}}_i}/\partial {\hat \tau _{i}}$ is given by
\begin{equation}\label{eq:r_tau}
\partial {{\bf{r}}_i}/\partial {\hat \tau _{i}} = {{\bf{M}}_1}{{\bf{\Delta }}_{{\hat {\nu}_i}}}{{\bf{M}}_2}{{\bf{\Psi }}_{{{\hat \nu}_i}}}{{\bf{M}}_3}{\bf{\Lambda }}_{{\hat \tau _{i}}}^\prime {{\bf{M}}_4}{{\bf{\Lambda }}_{{{{{\hat \nu}_i}}}}}{\bf{m}},
\end{equation}
and $\partial {{\bf{r}}_i}/\partial {\hat \nu_{i}}$ is given by
\begin{equation}\label{eq:dr_df}
\begin{aligned}
\partial {{\bf{r}}_i}/\partial {{\hat \nu}_{i}}
&= \mathbf{M}_1 {\bf{\Delta }}_{{{\hat \nu}_i}}^\prime\mathbf{M}_2 \mathbf{\Psi}_{{{\hat \nu}_i}} \mathbf{M}_3
\mathbf{\Lambda}_{\hat \tau_{i}}\mathbf{M}_4 \mathbf{\Lambda}_{\hat \nu_{i}} \mathbf{m} \\
&\quad + \mathbf{M}_1 {{\bf{\Delta }}_{{\hat \nu _i}}}\mathbf{M}_2 \mathbf{\Psi}_{{{\hat \nu}_i}}^{\prime} \mathbf{M}_3
\mathbf{\Lambda}_{\hat \tau_{i}}\mathbf{M}_4 \mathbf{\Lambda}_{\hat \nu_{i}} \mathbf{m} \\
&\quad + \mathbf{M}_1 {{\bf{\Delta }}_{{\hat \nu _i}}}\mathbf{M}_2 \mathbf{\Psi}_{{{\hat \nu}_i}} \mathbf{M}_3
\mathbf{\Lambda}_{\hat \tau_{i}}\mathbf{M}_4 \mathbf{\Lambda}_{\hat \nu_{i}}^{\prime} \mathbf{m},
\end{aligned}
\end{equation}
where 
\begin{equation}\label{eq:partial_matrix}
\begin{aligned}
\mathbf{\Lambda}'_{\hat \tau_{i}}
&= {\mathop{\rm diag}\nolimits} \left( { - {\rm{j}}2\pi \frac{m}{N}{\mkern 1mu} {e^{ - {\rm{j}}2\pi {\hat \tau _i}m/N}}} \right),\\
\mathbf{\Delta}'_{\hat \nu_i}
&= {\mathop{\rm diag}\nolimits} \left( {\left\{ {{\rm{j}}2\pi \frac{n}{N}{\mkern 1mu} {e^{{\rm{j}}2\pi {{\hat \nu}_i}n/N}}} \right\}_{n = 0}^{N - 1}} \right),\\
\mathbf{\Lambda}'_{\hat \nu_{i}}
&= {\mathop{\rm diag}\nolimits} \left( {{\rm{j}}2\pi \frac{p}{{NL}}{\mkern 1mu} {e^{{\rm{j}}2\pi {\hat \nu_i}p/(NL)}}} \right),\\
\mathbf{\Psi}'_{{{\hat \nu}_i}}
&= \operatorname{diag}\!\big(\varepsilon_i[0],\ldots,\varepsilon_i[N-1]\big),
\end{aligned}
\end{equation}
with $m,p \in \left[ {0,0.5NL - 1} \right] \cup \left[ { - 0.5NL, - 1} \right]$, and 
\begin{equation}\label{eq:ebu}
\varepsilon_i[n]=
\begin{cases}
-\mathrm{j}2\pi\,\dfrac{n}{N}\,e^{-\mathrm{j}2\pi \hat \nu_{i}n/N},
& n\in\{0,\ldots,N/2\},\\[4pt]
\mathrm{j}2\pi\,e^{\mathrm{j}2\pi \hat \nu_{i}\left(1-\frac{n}{N}\right)}
\left(1-\dfrac{n}{N}\right).
& \text{otherwise}.
\end{cases}
\end{equation}
The second-order partial derivatives of ${\bf r}_i$ with respect to $(\hat\tau_i,\hat\nu_i)$ can be derived similarly via the chain rule and are omitted due to space limitations. Substituting \eqref{eq:r_tau}, \eqref{eq:dr_df}, and these second-order derivatives into (18)--(22) in \cite{NOMP-OFDM} yields the required second-order derivatives of $S$, which are used in \eqref{eq:Newton_iter} to iteratively update $(\hat\tau_i,\hat\nu_i)$. The gain $\hat\beta_i$ is then updated as
\begin{equation}\label{eq:beta_i}
{\hat \beta _i} = {{\bf{r}}_i}^{\rm{H}}\left( {{{\hat \tau }_{i}},{{\hat \nu}_{i}}} \right){{\bf{h}}_r}/\left\| {{{\bf{r}}_i}\left( {{{\hat \tau }_{i}},{{\hat \nu}_{i}}} \right)} \right\|_2^2.
\end{equation}
The number of iterations in the single-refinement stage is set to $N_{{\rm iter,s}}$.
\subsubsection{Cyclic-Refinement Stage}
If $K$ propagation paths are estimated, we collect the $K$ parameter sets as ${\cal P} = \left\{ {\left( {{{\hat \tau }_{i}},{{\hat \nu}_{i}},{{\hat \beta }_i}} \right),i = 1, \ldots ,K} \right\}$. The estimates are then refined cyclically. When refining the $n$-th path, we form the residual by subtracting the contributions of all other paths, i.e.,
\begin{equation}\label{eq:cyclic_refine}
{{\bf{h}}_{r,n}} = {{\bf{r}}_{{\rm{prac}}}} - \sum\limits_{i = 1,k \ne n}^K {{{\hat \beta }_i}} {{\bf{r}}_i}\left( {{{\hat \tau }_{i}},{{\hat \nu}_{i}}} \right).
\end{equation}
Using ${{\bf{h}}_{r,n}}$ in place of ${{\bf{h}}_r}$ in the single-refinement stage, we refine the $n$-th parameter set ${\cal P}_n=\left( {{{\hat \tau }_{n}},{{\hat \nu}_{n}},{{\hat \beta }_n}} \right)$, with ${N_{{\rm{iter,c}}}}$ iterations performed for each set ${\cal P}_n$ update. After all $K$ parameter sets have been updated, the resulting residual is given by
\begin{equation}\label{eq:cyclic_refine_total}
{{\bf{h}}_r}\left( {\cal P} \right) = {{\bf{r}}_{{\rm{prac}}}} - \sum\limits_{i = 1}^K {{{\hat \beta }_i}} {{\bf{r}}_i}\left( {{{\hat \tau }_{i}},{{\hat \nu}_{i}}} \right). 
\end{equation}
In the next execution of steps i)-iii), we replace ${{\bf{h}}_r}$ in the coarse-estimation stage with ${{\bf{h}}_r}\left( {\cal P} \right)$. The iterations terminate when 
${\left\| {{{\bf{h}}_r}\left( {\cal P} \right)} \right\|_2} < \epsilon $, with $\epsilon$ denoting the threshold. 

\subsection{Computational Complexity}
The complexity is dominated by the FFT-based fractional circular shifts of the precomputed ambiguity samples $\bar{\bf A}_g[p,u]$ in \eqref{eq:r_factorized}. Accounting for the iterations, the per-estimation complexity is
$O\!\left(N_{{\rm iter,s}}N_{{\rm iter,c}}N_{{\rm main},\tau}N\log N\right)$,
where $N_{{\rm main},\tau}$ denotes the mainlobe width of the delay slices of $\bar{\bf A}_g$ and typically $N_{{\rm main},\tau}\ll N$, while $N_{{\rm iter,s}}$ and $N_{{\rm iter,c}}$ are usually much smaller than $N$. In contrast, the SBL scheme in \cite{SBLtang} (without hierarchical Laplace priors) has complexity $O\!\left((N_\tau N_\nu)^3\right)$, where $N_\tau$ and $N_\nu$ are the delay and Doppler grid sizes; typically $N_\tau N_\nu\approx N$ without refinement, and the complexity increases rapidly as the grid is refined.

\section{Simulation Results}

In this section, we evaluate the proposed AFNC method via Monte Carlo simulations. The AFDM parameters are set to $N=128$, $2Nc_1=7$, $c_2=0$, $f_{\rm c}=24~\mathrm{GHz}$, and $\Delta f=60~\mathrm{kHz}$. An RRC pulse-shaping filter with oversampling factor $L=4$ is employed. We consider a three-tap channel selected from the 3GPP TDL-C model \cite{3gpp_tr38901}, with normalized delays $(0.6366,\,2.7105,\,4.6003)$ and relative powers $(0,\,-13.2,\,-13.9)~\mathrm{dB}$. The maximum velocity is set to $v_{\max}=500~\mathrm{km/h}$, and each tap velocity is independently drawn from $[-v_{\max},v_{\max}]$ in each run. The proposed method is compared with the SBL scheme in \cite{SBLtang} (without hierarchical Laplace priors), where the dictionary grid resolutions are $\delta\tau_{\rm SBL}=1$ and $\delta f_{\rm SBL}=0.1$.

\begin{figure}
	\centering
	\includegraphics[width=0.3\textwidth]{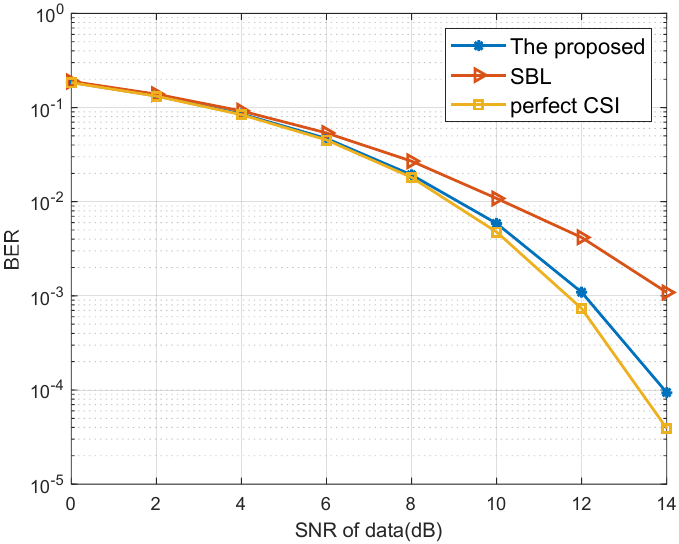}
	\caption{BER performance comparison of different algorithms at ${\rm{SN}}{{\rm{R}}_{\rm{p}}} = 30~\mathrm{dB}$.}
	\label{fg:BER_Newton_SBL}
	\vspace{-10pt}
\end{figure}

To evaluate the channel estimation performance, we use the normalized mean square error (NMSE), defined as
\begin{equation}\label{eq:NMSE}
{\rm{NMSE}} = \frac{{\left\| {{{\bf{H}}_{{\rm{es}}}} - {{\bf{H}}_{{\rm{true}}}}} \right\|_F^2}}{{\left\| {{{\bf{H}}_{{\rm{true}}}}} \right\|_F^2}},
\end{equation}
where ${{{\bf{H}}_{{\rm{true}}}}}$ and ${{{\bf{H}}_{{\rm{es}}}}}$ denote the true and the estimated channel matrices, respectively, and $\left\|  \cdot  \right\|_F^2$ denotes the squared Frobenius norm. 
\begin{figure}
	\centering
	\includegraphics[width=0.3\textwidth]{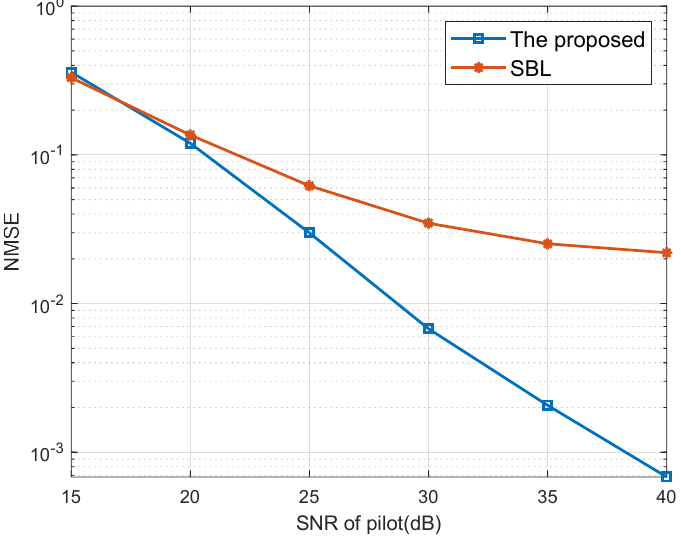}
	\caption{Channel estimation performance of the proposed method.}
	\label{fg:NMSE_Newton_SBL}
	\vspace{-10pt}
\end{figure}
Figure.~\ref{fg:NMSE_Newton_SBL} plots the NMSE of the proposed method and SBL versus the pilot SNR, defined as
${\rm SNR}_{\rm p}\triangleq P_{\rm pilot}/(P_{\rm n}/N)$,
where $P_{\rm pilot}$ is the pilot power and $P_{\rm n}$ is the total noise power over $N$ subcarriers. 
As ${\rm SNR}_{\rm p}$ increases, the proposed method achieves a lower NMSE by enabling off-grid super-resolution refinement, whereas SBL relies on an on-grid model.
Figure.~\ref{fg:BER_Newton_SBL} further evaluates the BER using the channel estimates at ${\rm SNR}_{\rm p}=30~\mathrm{dB}$ with minimum mean square error (MMSE) equalization and QPSK constellation, and include the perfect-CSI BER as a reference. The SNR of data is defined as ${\rm SNR}_{\rm d}\triangleq P_{\rm d}/P_{\rm n}$, with $P_{\rm d}$ denoting the power of data. The proposed method yields a significantly lower BER than SBL at ${\rm SNR}_{\rm p}=30~\mathrm{dB}$.

\section*{Acknowledgment}
This work was supported in part by the China Postdoctoral Science Foundation under Grant Number 2024M764088, in part by the National Natural Science Foundation of China under Grant 61971025 and 62331002.
\bibliographystyle{IEEEtran}
\bibliography{IEEEabrv,IEEE_RadarConf}

\end{document}